\documentclass[twocolumn,prl,aps,showpacs]{revtex4}
\usepackage{graphicx}

\newcommand{\be}{\begin{equation}}
\newcommand{\ee}{\end{equation}}
\newcommand{\bea}{\begin{eqnarray}}
\newcommand{\eea}{\end{eqnarray}}
\newcommand{\HH}{{\cal H}}

\newcommand{\p}{\partial}

\newcommand{\la}{\langle}
\newcommand{\ra}{\rangle}
\newcommand{\lb}{\left[}
\newcommand{\rb}{\right]}
\newcommand{\lp}{\left(}
\newcommand{\rp}{\right)}
\renewcommand{\vec}[1]{{\bf #1}}
\def\nn{\nonumber\\}

\begin{document}
\title{Coexistence of superfluid and Mott phases of lattice bosons}

\author{R.\,A. Barankov$^1$, C. Lannert$^2$ and S. Vishveshwara$^1$}
\affiliation{$^1$Department of Physics, University of Illinois at
Urbana-Champaign, 1110 W. Green St, Urbana, IL 61801, USA\\
$^2$Department of Physics, Wellesley College, Wellesley, MA 02481, USA}
\date{November 6th, 2006}
\begin{abstract}

Recent experiments on strongly-interacting bosons in optical lattices have
revealed the co-existence of spatially-separated Mott-insulating and
number-fluctuating phases.
The description of this
inhomogeneous situation is the topic of this Letter. We establish that the
number-fluctuating phase forms a superfluid trapped between the Mott-insulating
regions and derive the associated collective mode structure. We discuss the
interlayer's crossover between two- and three-dimensional behavior as a
function of the lattice parameters and estimate the critical temperatures for
the transition of the superfluid phase to a normal phase.

\end{abstract}  \keywords{}

\maketitle

Dilute gases of ultra-cold bosons on a lattice present a model system for
exploring quantum phases of matter. Experiments in optical lattice traps have
demonstrated controlled tunability through a quantum phase transition between a
Mott-insulating phase which has fixed particle number on each lattice site and
a phase exhibiting number fluctuations~\cite{Greiner02-1,Bloch05}. As has been
recently observed in radially symmetric traps~\cite{Foelling06,Campbell06}, for
sufficiently deep optical lattice potentials the system arranges itself into a
``wedding cake structure'' in which Mott-insulating phases of bosons
commensurate with the lattice alternate with interlayers of incommensurate
bosons with fluctuating site
occupancy~\cite{Jaksch98,Kashurnikov02,Wessel04,DeMarco05}. Various questions
concerning such inhomogeneous systems have so far been unanswered in both
theory and experiment: is the interlayer associated with number fluctuations a
condensate? If so, in what temperature range is the condensate robust? Is there
collective behavior in the interlayer, analogous to that seen in bulk
superfluids? How does the system cross-over from three-dimensional to
two-dimensional behavior as the interlayer thickness is varied? Understanding
these issues would also be important for related avenues in cold atomic
physics, such as the interplay of spatial inhomogeneity and quantum
criticality~\cite{Sachdev_book}, realization of robust states for quantum
computation~\cite{Cirac94}, and the physics of interacting fermions on a
lattice~\cite{Hofstetter02}.

It has been established that dilute gases of bosons in optical lattice
potentials are well-described by the Bose-Hubbard
Hamiltonian~\cite{Fisher89,Jaksch98} in which the bosons' movement between
sites is characterized by a tunneling term $\HH_J =-J\sum_{\la
ij\ra}a^\dag_ia_j$ related to the overlap of the single-particle wave functions
between neighboring sites $i$ and $j$, and the on-site interaction is modeled
by $\HH_U=(U/2)\sum_i n_i(n_i-1)$. It is the external trapping potential,
$V(\vec r_i)$, which is responsible for breaking the uniformity of the system
and promoting spatial coexistence of the Mott insulating and superfluid phases
at large interaction~\cite{Jaksch98,Kashurnikov02,Wessel04,DeMarco05}.
Analytical treatments of the inhomogeneous system are complicated by the fact
that no simple approximation of the Hamiltonian can faithfully describe the
entire phase space. The Bogoliubov approximation~\cite{Oosten02,Dalfovo99}
captures the condensed phase for large $J/U$ but breaks down close to the Mott
regions. The decoupled-site
approximation~\cite{Rokhsar91,Krauth92,Sachdev_book,Oosten02}, valid when $J
\ll U$ and the boson density is close to a commensurate value, works well
within and close to the Mott regions but fails deep within the incommensurate
phase. A third possibility is presented by the ``pseudo-spin"
approximations~\cite{Matsubara56,Bruder93,Altman02}, valid for intermediate
values of $J/U$, which bypass these shortcomings by treating kinetic energy and
interactions on comparable footing.

In this Letter, we employ a pseudo-spin approximation of the Bose-Hubbard model
to describe the inhomogeneous systems where the density of bosons varies as a
result of a confining trap. Concentrating on a single interlayer trapped
between two Mott-insulating phases, we show that number fluctuations give rise
to a condensate with a well-defined order parameter and derive the dynamical
equations governing the system. We obtain the collective excitation spectrum of
the interlayer condensate and show that in the homogeneous limit, it properly
reproduces the known properties of bulk superfluids. We explore the behavior of
the collective modes as a function of the thickness of the interlayer and show
that they provide a signature of dimensional cross-over in the condensate,
which can be achieved by tuning experimental parameters. We conclude with a
brief discussion of the expected mean-field critical temperature $T_c$ of an
interlayer superfluid and its behavior as a function of interlayer thickness.

Focusing our attention on the Mott phases with integer boson filling $n$ and
$n+1$ and the superfluid phase at intermediate fillings, we consider a Hilbert
space restricted to the number-basis states $|n\ra$ and $|n+1\ra$ at each site.
Considering the excluded states $|n-1\ra$, and $|n+2\ra$, we find that their
contribution to the energy is of order of $J^2/U$. We note that the
number-fluctuations on sites are driven by the incommensurability of bosons
with the lattice in the presence of the trapping potential. The truncated
Hilbert space in the limit $J/U\ll 1$ may be represented by the spin-$1/2$
states~\cite{Rokhsar91,Bruder93}, $|n+1\ra=|\uparrow\ra$ and
$|n\ra=|\downarrow\ra$, the eigenstates of the operator $s^z$ with eigenvalues
$\pm 1/2$. The tunneling term in the Bose-Hubbard Hamiltonian can be identified
with raising and lowering spin-$1/2$ operators, $s^+$ and $s^-$, such that $
a^\dagger_i a_j\to (n+1)s_i^+ s_j^-$, where $ a^\dagger_i$ is the boson
creation operator on site $i$. The interaction and the potential energy terms
are diagonal in the number basis at each site and the boson number operator
($\hat{n}_i = a_i^{\dag}a_i$) can be expressed in terms of the spin-1/2 matrix
$s^z$, $\hat n=n+1/2+s^z$. Thus, in the truncated Hilbert space, one obtains an
effective Hamiltonian identical to the spin-$1/2$ $XY$ model in the external
``magnetic" field:
\be\label{eq:Ham_spin}
\HH=-J(n+1)\sum_{\la ij\ra}\lp s^x_i s^x_j+s^y_i s^y_j\rp+\sum_i
(Un-\mu_i)s^z_i.
\ee
Here, $\la ij\ra$ denotes a summation over nearest neighbors sites, and
$\mu_i=\mu-V(\vec r_i)$ defines the chemical potential offset by the external
trapping potential, $V(\vec r_i)$. The chemical potential $\mu$ is set by the
total number of particles in the system, $\la N\ra=\sum_i \la \hat{n}_i \ra$.

\begin{figure}[t]
\includegraphics[width=3.in]{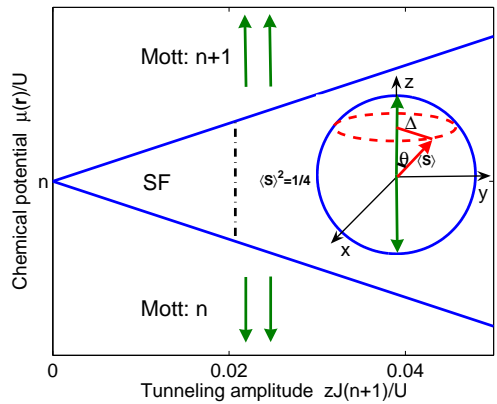}
%
\caption[]{The mean-field phase diagram for the Bose-Hubbard Hamiltonian. The
dash-dotted line corresponds to the interlayer with fluctuating site occupancy.
Spontaneous symmetry breaking in the ground state is shown on the sphere
$\la\vec s\ra^2=1/4$: the equilibrium configuration $\la \vec s\ra$ is
degenerate on the circle (dashed line) with nonzero order parameter
$|\Delta|=(1/2)\sin\theta$. North and South poles of the sphere correspond to
Mott states with $n+1$ and $n$ bosons per site, respectively.}
\label{fig:sphere}
\end{figure}

The pseudospin operators are coupled ferromagnetically in the $x$-$y$ plane and
therefore, at low temperatures can form an ordered state with broken $U(1)$
symmetry in the plane. At the mean-field level, in the ground state
configuration, pseudospins are aligned with the local ``magnetic" field, $\vec
B^0_i= z J(n+1)\lb 2\la s^x_i\ra,2\la s^y_i\ra,\cos\theta_i\rb$, where
$\cos\theta_i=(\mu_i-Un)/(zJ(n+1))$, and we have assumed $\la \mathbf{s}_i \ra
\approx \la \mathbf{s}_j\ra$ for nearest-neighbors. The equilibrium components
of the pseudospin at site $i$ are parameterized by angles on the sphere:
\be\label{eq:spin_ground}
\la s^z_i\ra=(1/2)\cos\theta_i,\quad \la
s^+_i\ra=(1/2)\,e^{i\varphi}\sin\theta_i,
\ee
where the angle $\varphi$ independent of site index expresses the phase
coherence in the system. The continuous degeneracy in the ground state is
illustrated in Fig.~\ref{fig:sphere}. In the Mott phase, the pseudospins are
completely polarized along the $z$ direction, i.e. $\la s^z_i\ra=\pm 1/2$,
allowing the identification of
 $\mu_{\pm}=Un\pm z J(n+1)$, the values of the chemical potential
at the boundaries of the Mott states with $n$ and $n+1$ bosons per site~(see
Fig.~\ref{fig:sphere}). In the $xy$-symmetry broken phase, $\la a^{\dag} \ra =
\la s^{+} \ra/\sqrt{n+1} \neq 0$ and we have a condensate with order parameter
\be\label{eq:net_order}
\Delta=(1/N_{\mbox{\scriptsize inter}})\sum_i\la s_i^+\ra,
\ee
where $N_{\mbox{\scriptsize inter}}$ is the number of lattice sites between the
two Mott phases.

The locations and sizes of the interlayers can be determined by the
relationship $\mu_{\pm}=\mu-V(\vec r_{\pm})$, where the chemical potential,
$\mu$, is obtained self-consistently by fixing the total number of particles in
the system, $N$. For radially-symmetric traps, a simplification occurs in the
limit of a thin interlayer, $\delta r_n\ll r_n$ (where $r_n$ is the radius at
the center of the interlayer between two Mott states with particle occupation
$n$ and $n+1$ and $\delta r_n$ is its thickness). In this case, the trapping
potential can be linearized around $r_n$ and we find that the number of
particles in the interlayer is the same as in the case $J/U=0$, where the
interlayer region would be filled with $n$ and $n+1$ Mott phases. Hence,  the
chemical potential at small $J/U$ can be found by setting
$N=(4\pi/3)\sum_{n=0}^{m-1}(r_n/\ell)^3$, where $m$ is the total number of Mott
states in the trap and $\ell$ is the lattice spacing~\cite{DeMarco05}. We find
that for a three-dimensional parabolic trapping potential, $V(r)=\alpha r^2$,
the interlayer parameters are given by $r_n=\lp\mu/\alpha\rp^{1/2}\lb
1-nU/\mu\rb^{1/2}$, $\delta r_n=6 J(n+1)/(\alpha r_n)$ when $n>0$, and $\delta
r_0=3 J/(\alpha r_0)$. These results show that it is possible to tune the width
of the interlayers from $\delta r_n\simeq \ell$ to $\delta r_n\gg \ell$,
effectively changing the dimensionality of the layers. As a characteristic
example in the range of recent experiments~\cite{Campbell06,Foelling06}, a
system with trap curvature $\alpha\approx h\times 24{\rm\,Hz/\mu m^2}$, total
particle number $N\approx 10^6$,
lattice spacing
$0.43{\rm\,\mu m}$, interparticle interaction $U\approx h\times 10{\rm\,kHz}$
and tunneling strength $J\approx h\times 120{\rm\,Hz}$ hosts two Mott regions
with $n=1$ and $n=2$, and two interlayers. The corresponding interlayer
parameters are $r_0\approx 25{\rm\,\mu m}$ and $\delta r_0 \approx 0.5{\rm\,\mu
m}$; $r_1\approx 14{\rm\,\mu m}$ and $\delta r_1 \approx 4{\rm\,\mu m}$. We
remark that for a harmonic trap, when gravity is taken into account, the system
experiences a shift along the direction of the gravitational field, but is
otherwise unaffected.

Having identified the interlayer region, we now turn to the low-energy
collective modes within the interlayer. These modes can be calculated using the
Heisenberg equations of motion for the pseudospin operators, $\p_t\vec
s_i=i[\HH,\vec s_i]$. In the mean-field approximation, one obtains the Bloch
equations, $\p_t\la\vec s_i\ra=\la\vec s_i\ra\times \vec B_i$, where the
effective magnetic field is given by $B^+_i=J(n+1) \sum_j 2\la s^+_j\ra$
(summation is over the nearest neighbors of site $i$), and
$B^z_i=zJ(n+1)\cos\theta_i$. Assuming that the characteristic wavelength of the
excitations is much larger than the lattice spacing $\ell$, we approximate the
sums entering the effective magnetic field $\vec B$ by their continuum limit,
$\sum_j \la\vec s_j\ra\approx z \la\vec s\ra+\ell^2\nabla^2\la\vec s\ra$. The
resulting Bloch equations can be analyzed as follows. The equilibrium number
density of bosons in the interlayer is $\rho_0=n+1/2 +\cos\theta/2$ and the
number density $\rho=\rho_0+\delta\rho$ obeys the continuity equation
$\p_t\rho+\nabla \vec j=0$ with the current density $\vec
j=(J(n+1)/2)\sin^2\theta\,\nabla\varphi$. Using the relationship between the
canonically conjugate density deviation $\delta\rho$ and the phase,
$\p_t\varphi=-2zJ(n+1)\delta\rho$, one obtains the following differential
equation for density fluctuations around equilibrium:
\be\label{eq:density_equation}
\p^2_{t}\,\delta \rho=4z\lp J(n+1)\ell\rp^2\nabla\lb\Delta_0^2\,\nabla \delta
\rho\rb,
\ee
where $\Delta_0=(1/2)\sin\theta$ is the local value of the order parameter
vanishing at the boundaries of the Mott states. The form of
Eq.(\ref{eq:density_equation}) is identical to that governing a trapped
Bose-Einstein condensate in the absence of a lattice~\cite{Pethick_Smith} with
$\Delta_0^2$ playing the role of an equilibrium density of the condensate
confined between two Mott states to an interlayer with radius $r_n$ and width
$\delta r_n$. It must be noted that while the equations governing density
distortions are identical to those derived from the standard Gross-Pitaevskii
formalism~\cite{Pethick_Smith} for a condensate in the absence of a lattice,
the equations of motion for the order parameter $\la s^+\ra$ in general do not
correspond to the Gross-Pitaevskii form, but reproduce it in the limit of small
density distortions.

For the uniform case, the excitation spectrum can be obtained by treating the
order parameter $\Delta_0$ as spatially-independent. The eigenvalue equation,
Eq.~(\ref{eq:density_equation}), is solved by the Fourier transformation,
$\delta \rho\propto \exp(i\vec p\vec r-i\omega_\vec p t)$, where $\vec p$ is
the wave vector. The resulting sound mode,
\be\label{eq:sound_mode}
\omega_\vec p=cp,\quad c=\sqrt{z}J(n+1)\ell |\sin\theta|,
\ee
is related to the spontaneously-broken symmetry in the ordered state. According
to the Landau criterion, the sound-like spectrum of Eq.~(\ref{eq:sound_mode})
makes the ordered state a superfluid. One notices that the speed of sound, $c$,
goes to zero as one approaches the Mott phases at $\sin\theta=0$.

In the trapped geometry, an estimate of the excitation spectrum can be obtained
from the quantization conditions imposed on the wave vector $p$ in
Eq.(\ref{eq:sound_mode}), with the speed of sound approximated by its value in
the center of the layer, $c_0=\sqrt{z}J(n+1)\ell$. For a spherically symmetric
trap, the excitation modes are confined within an interlayer centered at radius
$r_n$ with width $\delta r_n=2a_n$. The excitation modes terminate at the
boundaries of the Mott regions, \emph{i.e.} $p_j=j/a_n$ with $j=0,1,\ldots$,
which gives a radial mode spectrum $\omega_j\simeq \Omega_r j$ with
characteristic frequency $\Omega_r=J(n+1)\ell/a_n$. The quantization of the
surface modes is related to the angular momentum $L=0,1,\ldots$ through
$p_L=L/r_n$ which leads to the spectrum $\omega_L\simeq \Omega_a L$ with
characteristic frequency $\Omega_a=\Omega_r a_n/r_n$. The degeneracy of the
surface modes is $(2L+1)$ for each value of $L$. The perturbative calculation
of the modes in Eq.(\ref{eq:density_equation}) in the limit $a_n/r_n\ll 1$
confirms these estimates and gives the following result:
\bea\label{eq:spectrum_trap}
\lp\frac{\omega_{Lj}}{\Omega}\rp^2&\approx&j(j+1)+\frac{a_n^2}{r_n^2}\lb
1+\frac{3}{(2j-1)(2j+3)}\rb+\nn
&&L(L+1)\frac{a_n^2}{2r_n^2}\lb 1-\frac{1}{(2j-1)(2j+3)}\rb,
\eea
where $\Omega=\sqrt{6}\,\Omega_r$, $j=0,1,\ldots$, $L=0, 1,\ldots$, and $j+L\ne
0$. In the continuum approximation, the wavelength of the modes should be much
larger than the lattice spacing, which sets upper bounds on the quantum
numbers: $L\ll r_n/\ell$ and $j\ll a_n/\ell$. The second term in
Eq.(\ref{eq:spectrum_trap}) is independent of $L$ and is associated with the
curvature of the interlayer; it vanishes at $j=0$. The lowest energy modes for
thin interlayers, $a_n/r_n\ll 1$, correspond to angular excitations ($j=0$)
given by $\omega_L=2\Omega_a\sqrt{L(L+1)}$, $L=1,2,\dots$.

We note that the mode spectrum of Eq.~(\ref{eq:spectrum_trap}) corresponds to
that of a condensate confined with an explicitly shell-shaped trap (for
instance, a ``bubble trap" in Ref.~\cite{Zobay04}) since the ``effective
confining potential" in Eq.~(\ref{eq:density_equation}) has the form
$V_{eff}\propto (r-r_n)^2/a_n^2$ for thin interlayers. The calculation of the
radial ($L=0$) modes with $j=1,2$ in Ref.~\cite{Lannert06} confirms this
connection for the lowest-lying radial modes (analogous to ``breathers" in
spherical condensates).

The characteristic frequencies $\Omega_r$, $\Omega_a$ of the radial and angular
modes set temperature scales at which the spectrum in
Eq.(\ref{eq:spectrum_trap}) becomes quasiclassical, $j,L\gg 1$. For the
aforementioned experimental parameters, the corresponding energy scales are of
the order $\Omega_r\simeq 5{\rm\, nK}$, $\Omega_a\simeq 0.5{\rm\,nK}$. The
energy of the system at finite temperature is obtained through quantization of
the collective modes, $E(T)=\sum_{Lj}(2L+1)\omega_{Lj}n_{Lj}$, where
$n_{Lj}=1/(\exp(\omega_{Lj}/T)-1)$ is the thermal occupation of the bosonic
modes with spectrum given by Eq.~(\ref{eq:spectrum_trap}), and the factor
$(2L+1)$ takes into account the degeneracy of the angular modes. There are
three temperature regimes in this case. In the extreme low-temperature limit,
$T\ll\Omega_a$, thermal excitations are gapped, i.e. $E(T)\propto
\Omega_a\exp(-2\sqrt{2}\Omega_a/T)$. At intermediate temperatures, $\Omega_a\ll
T\ll\Omega_r$, the radial modes are frozen and only the two-dimensional angular
modes contribute to the energy, $E(T)\propto T^3/\Omega_a^2$. At higher
temperatures, $T\gg\Omega_r$, both radial and angular modes are excited, and
the energy has a three-dimensional phonon-like temperature dependence,
$E(T)\propto T^4/\Omega_r^3$. The separation of temperature scales achieved by
changing the interlayer width from $\delta r_n\simeq \ell$ to $\delta r_n\gg
\ell$ tunes the effective dimensionality of the system from two to three
dimensions.

At finite temperatures, the order parameter introduced in
Eq.(\ref{eq:net_order}) is depleted from its zero temperature value by the
collective modes. In the low-temperature regime, $T\ll J(n+1)$, for wide
interlayers, $\delta r_n\gg \ell$, the order parameter depletion is similar to
the case of a three-dimensional weakly-interacting BEC, $\delta
\Delta(T)\propto (T/J(n+1))^2$. At higher temperatures, the long-range order is
destroyed by the quasiparticle excitations whose wavelength is of the order of
the lattice spacing. The critical temperature ought to be of the same order of
magnitude as these excitations with the energies of order $J(n+1)$ (obtained
from setting $j\simeq a_n/\ell$ and $L\simeq r_n/\ell$ in
Eq.(\ref{eq:spectrum_trap})). A mean-field calculation similar to the one in
Ref.\cite{Matsubara56} confirms the estimate and provides a mean-field
expression for the critical temperature $T_{3D}=3J(n+1)$. Thermal properties of
narrow interlayers, $\delta r_n\simeq \ell$, are qualitatively different. In
this case the angular excitations play the dominant role. In accordance with
the Mermin-Wagner-Hohenberg theorem~\cite{MWH-theorem}, the long-range order is
destroyed but the system retains power-law correlations in the phase of the
order parameter. In the limit that the interlayer width is comparable to the
lattice spacing, $\delta r_n\simeq \ell$, a simple model capturing the
properties of the two-dimensional system involves only phase variables and
leads to the effective Hamiltonian, $\HH_{\varphi}=(K/2)\int d^2 x\lp
\nabla\varphi\rp^2$, where the integration is over the surface of the spherical
layer, and we have introduced the phase stiffness $K=J(n+1)/2$.  The
Kosterlitz-Thouless (K-T) transition~\cite{Kosterlitz73} between the
high-temperature normal and the low-temperature superfluid state occurs at
temperature $T_{2D}=(\pi/2)K=(\pi/4)J(n+1)$. At intermediate widths, $\delta
r_n\gtrsim \ell$, the phase stiffness is approximated by $K=(J(n+1)/2)(\delta
r_n/\ell)\,\overline{\sin^2\theta}$ with $\overline{\sin^2\theta}=(1/\delta
r_n)\int dr\sin^2\theta$. For the interlayer in the trap
$\overline{\sin^2\theta}=2/3$, and the critical temperature of the K-T
transition is given by $T_c=(\pi/6)(\delta r_n/\ell)J(n+1)$, which is a linear
function of the interlayer width, interpolating between two-dimensional and
three-dimensional limits, $T_{2D}\le T_c\le T_{3D}$. In the range of current
experiments, for $J\approx h\times 120{\rm Hz}$, one obtains an estimate of the
critical temperature, $T_c\simeq 10{\rm\, nK}$.

In conclusion, we have shown that the interlayer with fluctuating site
occupation confined between two Mott states becomes superfluid at low but
experimentally accessible temperatures. Employing the pseudospin model, we have
identified the effective potential confining the superfluid and analyzed the
low-energy excitations in the system. We have demonstrated that the effective
dimensionality of the interlayer can be changed by tuning external parameters.
As an example of the ensuing physics we have suggested that the critical
temperature interpolates between two-dimensional and three-dimensional limits
as one changes the width of the interlayer. A clear experimental signature of
the interlayer condensate, either through time-of-flight and interference
experiments, excitation of collective modes or radio-frequency spectroscopy is
yet to be obtained.

The authors are grateful for their discussions with A.~J.~Leggett and
A.~Auerbach. One of the authors (C.~L.) acknowledges support by the NSF through
grant DMR-0605871.


\begin{thebibliography}{10}

\bibitem{Greiner02-1}M. Greiner, O. Mandel, T. Esslinger, T. W. H\"ansch, and
I. Bloch, Nature {\bf 415}, 39 (2002).

\bibitem{Bloch05} I. Bloch, Nature Physics {\bf 1}, 23 (2005).

\bibitem{Foelling06}S. F\"olling, A. Widera, T. M\"uller, F. Gerbier, and I. Bloch,
Phys. Rev. Lett. {\bf 97}, 060403 (2006).

\bibitem{Campbell06}G. K. Campbell, J. Mun, M. Boyd, P. Medley, A. E. Leanhardt,
L. Marcassa, D. E. Pritchard, and W. Ketterle, Science {\bf 313}, 649 (2006).

\bibitem{Jaksch98}D. Jaksch, C. Bruder, J. I. Cirac, C. W. Gardiner, and P. Zoller,
Phys. Rev. Lett. {\bf 81}, 3108 (1998).

\bibitem{Kashurnikov02} V. A. Kashurnikov, N. V. Prokof'ev, and B. V. Svistunov,
Phys. Rev. A {\bf 66}, 031601(R) (2002).

\bibitem{Wessel04} S. Wessel, F. Alet, M. Troyer, and G. G. Batrouni,
Phys. Rev. A {\bf 70}, 053615 (2004).

\bibitem{DeMarco05}B. DeMarco, C. Lannert, S. Vishveshwara, and T. C. Wei,
Phys. Rev. A {\bf 71}, 063601 (2005).

\bibitem{Sachdev_book}S. Sachdev, {\it Quantum Phase Transitions}, Cambridge
University Press, 1999.

\bibitem{Cirac94}J. I. Cirac and P. Zoller, Phys. Rev. Lett. {\bf 74}, 4091
(1995).

\bibitem{Hofstetter02}W. Hofstetter, J. I. Cirac, P. Zoller, E. Demler,
and M. D. Lukin, Phys. Rev. Lett. {\bf 89}, 220407 (2002).

\bibitem{Fisher89}M. P. A. Fisher, P. B. Weichman, G. Grinstein,
and D. S. Fisher, Phys. Rev. B {\bf 40}, 546 (1989).

\bibitem{Oosten02}D. van Oosten, P. van der Straten, and H. T. C. Stoof,
Phys. Rev. A {\bf 63}, 053601 (2002).

\bibitem{Dalfovo99}F. Dalfovo, S. Giorgini, L. P. Pitaevskii, and S. Stringari,
Rev. Mod. Phys. {\bf 71}, 463 (1999).

\bibitem{Rokhsar91}D. S. Rokhsar and B. G. Kotliar, Phys. Rev. B {\bf 44},
10328 (1991).

\bibitem{Krauth92}W. Krauth, M. Caffarel, and J. P. Bouchaud, Phys. Rev. B {\bf
45}, 3137 (1992).

\bibitem{Matsubara56}T. Matsubara and H. Matsuda, Progress of Theoretical Physics,
{\bf 16}, 569 (1956).

\bibitem{Bruder93}C. Bruder, R. Fazio, and G. Sch\"on, Phys. Rev. B {\bf 47},
342 (1993).

\bibitem{Altman02}E. Altman and A. Auerbach, Phys. Rev. Lett. {\bf 89}, 250404
(2002).

\bibitem{Pethick_Smith}C. J. Pethick and H. Smith,
{\it Bose-Einstein condensation in dilute gases}, Cambridge University Press,
2002.

\bibitem{Zobay04} O. Zobay and B. M. Garraway, Phys. Rev. Lett. {\bf 86},
1195 (2001); Phys. Rev. A {\bf 69}, 023605 (2004).

\bibitem{Lannert06} C. Lannert, T.-C. Wei, and S. Vishveshwara, cond-mat/0604455.

\bibitem{MWH-theorem}N. D. Mermin and H. Wagner, Phys. Rev. Lett. {\bf 22}, 1133
(1966); P. C. Hohenberg, Phys. Rev. {\bf 158}, 383 (1967).

\bibitem{Kosterlitz73}J. M. Kosterlitz and D. J. Thouless,
J. Phys. C {\bf 6}, 1181 (1973).
\end{thebibliography}
\end{document}